\documentstyle[12pt]{article}
\begin{document}
\parskip 10pt plus 1pt
\title{The CNOT Quantum Logic Gate Using q-Deformed Oscillators}
\author{
{\it Debashis Gangopadhyay}\\
{S.N.Bose National Centre For Basic Sciences}\\
{JD Block, Sector-III, Salt Lake, Kolkata-700098, INDIA}\\
{\it debashis@boson.bose.res.in}\\
}
\date{}
\maketitle
\baselineskip=20pt
\begin{abstract}

It is shown that the two qubit CNOT (controlled NOT) gate can also be 
realised using q-deformed angular momentum states constructed via the 
Jordan-Schwinger mechanism.Thus all the three gates necessary for 
universality i.e.Hadamard, Phase Shift and the two qubit CNOT gate are 
realisable with q-deformed oscillators.


\end{abstract}
\newpage

{\bf 1. Introduction}

Recently$^{1}$ it has been shown that the single qubit quantum logic gates,
{\it viz.} ,the Hadamard and Phase Shift gates can also be realised
with two q-deformed oscillators where $q$ is the deformation
parameter of a quantum Lie algebra$^{2}$. q-Deformed oscillators here mean 
that the Lie  algebra satisfied by creation ($a^{\dagger}$) and annihilation 
($a$) operators of a bosonic oscillator, {\it viz.} $aa^{\dagger}-a^{\dagger}a=1$ 
is modified into $a_{q}a_{q}^{\dagger}-qa_{q}^{\dagger}a_{q}=q^{-N}$ where $N$ 
is the number operator and $q=e^{s}\enskip;\enskip 0<s<1$. Using such deformed 
oscillators an alternative formalism for quantum computation can be set up$^{1}$.
The advantage of this over the conventional formalism (which is obtained for 
$q\rightarrow 1$) is the presence of an arbitrary function which may be
exploited for experimental purposes.
 
However, the formalism will be more meaningful if the realisation with
q-deformed qubits is possible for all the gates required for {\it universality}. 
A set of gates is said to be {\it universal for
quantum computation} if any unitary operation may be approximated to 
arbitrary accuracy by a quantum circuit involving those gates. In the case 
of standard quantum computation the Hadamard, Phase Shift and the CNOT 
(controlled NOT) gates constitute such a set$^{3}$. In this paper I show that 
the 2-qubit controlled-NOT gate can also be realised with q-deformed qubits.
Thus all the three gates, i.e. Hadamard, Phase Shift and CNOT gates
can now be obtained with q-deformed qubits. 

The motivation for considering q-deformed oscillators in quantum computation 
comes from the fact that deformed oscillators have been successfully used as a tool 
to understand deviations from an ideal theoretical or experimental 
scenario$^{4}$ for past many years. Bonatsos and Daskoloyamis$^{4}$ were among 
the firsts to show that the vibration spectra 
of diatomic molecules gave better fits using deformed oscillators.
Parisi$^{4}$ studied a d-dimensional array of Josephson junctions in a magnetic
field and computed the thermodynamic properties in the high temperature 
region for $d\rightarrow\infty$. Evaluation of the high temperature expansion
coefficients were done by mapping onto the computation of some matrix elements
for the q-deformed harmonic oscillator. Raychev$^{4}$ {\it et al} calculated the deviations  
from the nuclear shell model using the q-deformed  three dimensional
harmonic oscillator.Bonatsos, Lewis,Raychev and Terziev$^{4}$ demonstrated
that the three dimensional q-deformed harmonic oscillator correctly predicts
the first supershell closure in alkali clusters without introducing additional
parmeters.McEWan and Freer$^{4}$ showed that the nuclear orbitals of certain
nuclei were commensurate with the energy level scheme of the deformed harmonic
oscillator and the Nilsson model. For these reasons it is meaningful to study 
whether deformed oscillators can be used in the formalism of quantum computation.

A brief review of relevant facts is given in section 2. In Section 3 the NOT gate 
is realised with q-deformed oscillators. Section 4 gives the realisation of the 
two qubit CNOT gate in terms of q-deformed oscillators. In Section 5 the states 
are discused and Section 6 is the conclusion where a brief elaboration of
the possibility of  quantm error correction using deformed oscillators is given.  

{\bf 2.Brief Review}

Quantum logic gates are basically unitary operators $^{5-9}$.
Three gates,the single qubit  Hadamard and 
Phase Shift gates and the 2-qubit CNOT gate, are sufficient to construct any 
unitary operation on a single qubit $^{3}$.This is the {\it universality}
referred to above. These gates are constructed using the "spin up" and 
"spin down" states of $SU(2)$ angular momentum i.e., the two possible states 
of a qubit are  usually represented by "spin up" and "spin down" states.
This is the Jordan-Schwinger construction using two independent
harmonic oscillators. A similar construction 
of q-deformed angular momentum states can be done using two q-deformed 
oscillators$^{10}$. In Ref.1 it was shown that the Hadamard and Phase Shift gates can also be
realised with q-deformed qubits. To achieve this ,the technique of harmonic 
oscillator realisation $^{11,12}$ of q-oscillators was used. This allows one 
to set up an alternate quantum computation formalism. 

q-Oscillators are described by deformed creation and annihilation operators,  
$a_{q}^{\dagger}$,$a_{q}$ ,respectively.
For ordinary  oscillators these are $a^{\dagger}$ and $a$.
$q= e^{s}$, $0\leq s\leq 1$ and 
the deformed oscillators satisfy the following relations :
$$a_{q}a_{q}^{\dagger}- qa_{q}^{\dagger}a_{q}=q^{-N}~~ ;~~ N^{\dagger}=N\eqno(1a)$$
$$[N,a_{q}]=-a_{q}~;~ [N,a_{q}^{\dagger}]=a_{q}^{\dagger}~;~ a_{q}^{\dagger}a_{q} = [N];~
a_{q}a_{q}^{\dagger}=[N+1]\eqno(1b)$$
$$a_{q}f(N) = f(N+1) a_{q}~~;~~a_{q}^{\dagger}f(N)=f(N-1)a_{q}^{\dagger}\eqno(1c)$$
where the q-number $[x]=(q^{x}-q^{-x})/(q-q^{-1})$ becomes  the ordinary number $x$
when $q\rightarrow 1$ (i.e.$s\rightarrow 0$). $N$ is the  number operator for the
q-deformed oscillators and $f(N)$ is any function of $N$.
The eigenvalue $n$ of the number operator $N$ denotes the number of bosonic particles.
We confine to real $q$. $a_{q},a_{q}^{\dagger}$ and  $a,a^{\dagger}$ are related as $^{11}$
$$a_{q}=a\sqrt{{\frac {q^{\hat N} \psi_{1} - q^{-\hat N}\psi_{2}}{{\hat N (q-q^{-1})}}}}\enskip;\enskip
a_{q}^{\dagger}=\sqrt{{\frac {q^{\hat N} \psi_{1} - q^{-\hat N}\psi_{2}}{{\hat N (q-q^{-1})}}}} a^{\dagger}\eqno(2a)$$
$$ N = \hat N - (1/s) ln~ \psi_{2}\eqno(2b)$$
$\hat N$ is the number operator for usual oscillators with eigenvalue $\hat n$;
and $\psi_{1}$ , $\psi_{2}$ are arbitrary functions of $q$ only with $\psi_{1,2}(q)= 1$ for $q=1$.

If all these arbitrary functions are unity, then $N=\hat N$, i.e.
the number operator of the deformed oscillator becomes identical
to the number operator of the standard harmonic oscillator. But 
oscillator states are usually expressed in occupation number 
basis. So if the number operators are identical, there is no 
way of differentiating between a deformed oscillator state 
and a standard oscillator state. So nothing is gained and we are
still in the realm of standard quantum computation.    
But equations $(2a,b)$ are general if the
$\psi_{i}(q),i=1,2$ are {\it not all equal to unity}. Let 
$\psi_{1}=\psi_{2}=\psi(q)$ . Now $N=\hat N - (1/s)~ln~ \psi(q)$ (equation $(2b)$). 
This will be reflected in the Jordan-Schwinger construction
of angular momentum states and the states in the two cases will be 
distinguishable through the function $\psi(q)$. Further details are in Ref.[1].

We now express a single qubit state in terms of two harmonic oscilator states 
using the Jordan-Schwinger construction.

(a) States  are defined by the total angular momentum $j$ and
$z$-component of angular momentum  $j_{z}$ i.e. $m$. A particular $(j,m)$ state 
is created by acting the creation operators on the vacuum or ground state which
is a direct product state of the individual oscillator ground states :
$$\vert j m\rangle= {\frac {( a^{\dagger}_{1})^{j+m} (a^{\dagger}_{2})^{j-m}}{[(j+m)!(j-m)!]^{1/2}}}
\vert\phi\rangle\eqno(3)$$
$\vert \phi\rangle\equiv \vert\tilde 0\rangle = \vert\tilde 0\rangle_{1} \vert\tilde 0\rangle_{2}$ is
the ground state ($j=0,m=0$), while $\vert\tilde 0\rangle_{i}, i=1,2$ are the oscillator
ground states.
$j=(n_{1}+n_{2})/2\enskip;\enskip m=(n_{1}-n_{2})/2$ where $n_{1},n_{2}$ are the 
eigenvalues of the number operators of the two oscillators.\enskip\enskip\enskip

(b) A qubit can be either "up" or "down" i.e. there are two possible configurations.
So the oscillator number operators can take the following sets of values only :
($n_{1}=1,n_{2}=0$, "up" state) and ($n_{1}=0, n_{2}=1$, "down" state). Hence  
$(n_{1}+n_{2})/2 = 1/2$. As $j=1/2$ for both qubit states,  suppress $j$ for 
simplicity of notation : 
$$\vert  m\rangle={\frac {(a^{\dagger}_{1})^{1/2 + m} (a^{\dagger}_{2})^{1/2  - m}}
{[(1/2  + m)!(1/2  - m)!]^{1/2}}}\vert\phi\rangle\enskip;\enskip
\vert - m\rangle={\frac {(a^{\dagger}_{1})^{1/2 - m} (a^{\dagger}_{2})^{1/2  + m}}
{[(1/2  + m)!(1/2  - m)!]^{1/2}}}\vert\phi\rangle \eqno(4a)$$
Equivalently, in terms of $n_{1},n_{2}$ these are 
$$\vert n_{1} -  1/2\rangle ={\frac {(a^{\dagger}_{1})^{n_{1}} (a^{\dagger}_{2})^{1  - n_{1}}}
{[(n_{1})!(1 - n_{1})!]^{1/2}}}\vert\tilde 0\rangle\enskip;\enskip
\vert - (n_{1} -  1/2)\rangle ={\frac {(a^{\dagger}_{1})^{1-n_{1}} (a^{\dagger}_{2})^{n_{1}}}
{[(n_{1})!(1 - n_{1})!]^{1/2}}}\vert\tilde 0\rangle \eqno(4b)$$

(c)The basis states are
($\vert 1\rangle\equiv \vert up\rangle$ state and $\vert 0\rangle\equiv \vert down\rangle$ state )
$$\vert 1\rangle\equiv\vert 1/2,1/2\rangle
\equiv\vert 1/2\rangle=a^{\dagger}_{1}\vert \tilde 0\rangle
=a^{\dagger}_{1}\vert\tilde 0\rangle_{1}\vert\tilde 0\rangle_{2}
=\vert\tilde 1\rangle_{1}\vert\tilde 0\rangle_{2}$$
$$\vert 0\rangle\equiv\vert 1/2,-1/2\rangle
\equiv\vert -1/2\rangle= a^{\dagger}_{2}\vert \tilde 0\rangle
=a^{\dagger}_{2}\vert\tilde 0\rangle_{1}\vert\tilde 0\rangle_{2}
=\vert\tilde 0\rangle_{1}\vert\tilde 1\rangle_{2}$$
The{\it physical meaning} of the  notation is as follows.
The $\vert 1\rangle$ (spin "up") state is constructed out of
two oscillator states where the first oscillator state has occupation
number $1$ while the other has occupation number $0$. The $\vert 0\rangle$
( spin "down") state corresponds to the first oscillator having 
occupation number $0$ and the second oscillator having occupation number 
$1$. So any qubit state $\vert x\rangle$ is :
$$\vert x\rangle=(a_{1}^{\dagger})^{x} (a_{2}^{\dagger})^{1-x}\vert \tilde 0\rangle\eqno(5)$$
$\vert 0\rangle$ represents one of the two possible qubit states while 
$\vert\tilde 0\rangle$ represents oscillator ground state i.e. occupation number $0$;
$\vert\tilde 1\rangle$ represents an oscillator state with occupation number $1$ etc.

Now consider the Hadamard transformation.
For a standard qubit, the Hadamard gate acts as follows. When one of the 
basis states is given as an input, the output is a superposition
of the two basis states,i.e. $\vert 0\rangle\rightarrow \vert 0\rangle +\vert 1\rangle$ 
and $\vert 1\rangle\rightarrow\vert 0\rangle - \vert 1\rangle$.
So the Hadamard transformation on a single qubit state ($x=0,1$) can be represented as
(modulo $ 1/\sqrt {2}$)
$$\vert x\rangle\longrightarrow (-1)^{x}\vert x\rangle ~+~\vert 1-x\rangle\eqno(6)$$
i.e.
$$\vert n_{1} - 1/2\rangle\longrightarrow (-1)^{n_{1}}\vert n_{1} - 1/2\rangle ~+~ \vert 1/2  -  n_{1}\rangle\eqno(7)$$
Following the discussion preceding equation $(3)$, the general q-deformed state is  
$\vert j m\rangle_{q} \equiv {\frac {(a_{1q}^{\dagger})^{n_{1}} (a_{2q}^{\dagger})^{n_{2}}}{([n_{1}]![n_{2}]!)^{1/2}}} 
\vert\phi\rangle_{q}$;
$\vert j ~~~-m\rangle_{q} \equiv {\frac {(a_{1q}^{\dagger})^{n_{2}} (a_{2q}^{\dagger})^{n_{1}}} 
 {([n_{1}]![n_{2}]!)^{1/2}}} \vert\phi\rangle_{q}$
where $\vert \phi\rangle_{q}\equiv \vert\tilde 0\rangle_{q}=\vert\tilde 0\rangle_{1q}\vert\tilde 0\rangle_{2q}$ 
is the ground state corresponding to two non-interacting
q-deformed oscillators $^{10}$. In our notation a qubit  state has either  (a) $n_{1}=0 , n_{2}=1$ or 
(b) $n_{1}=1 , n_{2}=0$. Hence
$$\vert n_{1} -  1/2\rangle _{q}
\equiv {\frac {(a_{1q}^{\dagger})^{n_{1}} (a_{2q}^{\dagger})^{1-n_{1}}} 
 {([n_{1}]![1-n_{1}]!)^{1/2}}} \vert \tilde 0\rangle_{q}\enskip;\enskip
\vert - (n_{1} -  1/2)\rangle_{q}
\equiv {\frac {(a_{1q}^{\dagger})^{1-n_{1}} (a_{2q}^{\dagger})^{n_{1}}} 
 {([n_{1}]![1-n_{1}]!)^{1/2}}}\vert\tilde 0\rangle_{q}\eqno(8)$$
So the Hadamard transformation for q-deformed state is 
$$\vert n_{1} - 1/2\rangle_{q} \longrightarrow (-1)^{n_{1}} 
\vert n_{1} - 1/2\rangle_{q} +\vert 1/2  - n_{1}\rangle_{q}\eqno(9)$$
This simplifies to:
$$[F_{1}(\hat N_{1},q)a_{1}^{\dagger}]^{n_{1}} [F_{2}(\hat N_{2},q)a_{2}^{\dagger}]^{1-n_{1}}
\vert\phi\rangle_{q}\longrightarrow$$ 
$$(-1)^{n_{1}}[F_{1}(\hat N_{1},q)a_{1}^{\dagger}]^{n_{1}}
[F_{2}(\hat N_{2},q)a_{2}^{\dagger}]^{1-n_{1}}\vert\phi\rangle_{q}$$
$$+ [F_{1}(\hat N_{1},q)a_{1}^{\dagger}]^{1-n_{1}} [F_{2}(N_{2},q)a_{2}^{\dagger}]^{n_{1}} 
\vert\phi\rangle_{q}\eqno(10)$$
where
$$F_{1}(\hat N_{1}, q)
=\sqrt{{\frac {q^{\hat N_{1}} \psi_{1} - q^{-\hat N_{1}}\psi_{2}} {{\hat N_{1} (q-q^{-1})}}}}\enskip,\enskip
F_{2}(\hat N_{2}, q)
=\sqrt{{\frac {q^{\hat N_{2}} \psi_{3} - q^{-\hat N_{2}}\psi_{4}} {{\hat N_{2} (q-q^{-1})}}}}
\eqno(11)$$
$n_{1},n_{2}$ is always $0$ or $1$ so as to correspond to the qubit. 
It is simple to check that the q-number $[0]$ is equal to the ordinary
number $0$ and simlarly the q-number $[1]$ equals odinary number $1$.
Hence the q-numbers 
$[n_{1}], [n_{2}]$ are always the usual numbers $n_{1}, n_{2}$.
Same restrictions also apply to usual (i.e.undeformed)  oscillators.
So we restrict the hatted number operators, 
$\hat N_{1}$ and $\hat N_{2}$,  by $\hat N_{1} + \hat N_{2} = I$ where $I$ is 
the identity operator.

The Phase Shift transformation of qubit states is :
$\vert x\rangle\longrightarrow e^{i x \theta}\vert x\rangle$ 
i.e $\vert 0\rangle\rightarrow \vert 0\rangle$ and $\vert 1\rangle\rightarrow e^{i\theta}\vert 1\rangle$.
which in our notation is
$\vert n-{\frac 12}\rangle\longrightarrow e^{in\theta}\vert n - {\frac 12}\rangle$
where $\theta$ is the phase shift. Now one can proceed as described in the previous sections. 
Details are in Ref.[1]. There it was shown that both the Hadamard and Phase Shift transformations 
can be realised with q-deformed qubits. Below it is shown that the same is possible for 
both the NOT gate and the CNOT  gate.

{\bf 3.The NOT gate}

The NOT gate essentially flips a qubit, i.e. $\vert 0\rangle\rightarrow\vert 1\rangle$ and 
$\vert 1\rangle\rightarrow\vert 0\rangle$. It acts on a qubit as :
$\vert x\rangle\rightarrow \vert 1- x\rangle$ where $x=0,1$. In our notation this is
$\vert n_{1}-{1\over 2}\rangle\rightarrow\vert {1\over 2}-n_{1}\rangle$.
For q-deformed states this means 
$\vert n_{1}-{1\over 2}\rangle_{q}\rightarrow\vert {1\over 2}-n_{1}\rangle_{q}$.
In terms of q-deformed oscillator states this becomes
$${\frac {(a_{1q}^{\dagger})^{n_{1}} (a_{2q}^{\dagger})^{n_{2}}} {{([n_{1}]![n_{2}]!)^{1/2}}}}
\vert\phi\rangle_{q}\rightarrow                                   
{\frac {(a_{1q}^{\dagger})^{n_{2}} (a_{2q}^{\dagger})^{n_{1}}} {([n_{2}]![n_{1}]!)^{1/2}}}
\vert\phi\rangle_{q}\eqno(12)$$
i.e. the exponents of the two creation operators are interchanged. Rewritten in terms of 
the functions $F$ this looks like
$$[F(\hat N)a_{1}^{\dagger}]^{n}[F(1-\hat N)a_{2}^{\dagger}]^{1-n}\vert\phi\rangle_{q}
\rightarrow 
[F(\hat N)a_{1}^{\dagger}]^{1-n}[F(1-\hat N)a_{2}^{\dagger}]^{n}\vert\phi\rangle_{q}$$
where one has used the fact that $n_{1}+n_{2}=1$ and followed the arguments 
{\it after equation $(13b)$ of Ref.1}, relabelled $n_{1}$ as $n$ etc.
Using $(1c)$ one gets
$$[F(\hat N)]^{n}[F(1+n -\hat N)]^{1-n}(a_{1}^{\dagger})^{n}(a_{2}^{\dagger})^{1-n}\vert\phi\rangle_{q}$$
$$\rightarrow 
[F(\hat N)]^{1-n}[F(2-n-\hat N)]^{n}(a_{1}^{\dagger})^{1-n}(a_{2}^{\dagger})^{n}\vert\phi\rangle_{q}\eqno(13)$$
With respect to the states $\vert\phi\rangle_{q}$, the above expression would be indistinguishable from 
the usual "NOT" transformation if 
$$[F(\hat N)]^{n}[F(1+n -\hat N)]^{1-n}=
[F(\hat N)]^{1-n}[F(2-n-\hat N)]^{n}$$
which simplifies to 
$$ F(\hat N)=F(1-\hat N)\eqno(14)$$
for both $n=0$ and $n=1$.

Written in terms of its eigenvalues means
$${\frac {\psi_{1}(q)}{\psi_{2}(q)}}={\frac {(q^{-\hat n}-\hat n q^{-\hat n}-\hat n q^{\hat n-1})}{(q^{\hat n}
-\hat n q^{\hat n}-\hat n q^{1-\hat n})}}
\eqno(15)$$
This has the solution $\psi_{1}(q)=\psi_{2}(q)=\psi(q)$(say) for both $\hat n=0$ and $\hat n=1$. 
Thus the NOT gate is realisable with deformed qubits.Moreover, the conditions
for realisation is the same (i.e. $\psi_{1}(q)=\psi_{2}(q)=\psi(q)$)
as for the Hadamard and Phase Shift gates.

\newpage

{\bf 4. The CNOT gate}

The Controlled-NOT gate is a two-qubit operator where the first qubit is the control and the second qubit
the target. The action of the  CNOT gate is defined by the following transformations:
$$\vert 00\rangle\rightarrow \vert 00\rangle \enskip; \enskip\vert 01\rangle\rightarrow \vert 01\rangle$$
$$\vert 10\rangle\rightarrow \vert 11\rangle\enskip; \enskip\vert 11\rangle\rightarrow \vert 10\rangle$$
where $\vert 00\rangle\equiv \vert 0\rangle\vert 0\rangle; \vert 01\rangle\equiv \vert 0\rangle\vert 1\rangle$ etc.
The first line of the transformation signifies that when the control qubit is in the $"0"$-state , the target 
qubit does not change after the action of the CNOT gate. The second line means that if the control qubit is in 
the $"1"$-state, target qubit changes value after the action of the CNOT gate. 
This may be written as (modulo constants) as
$\vert x y\rangle\rightarrow (1-x)\vert x y\rangle + x\vert x\enskip\enskip  1-y\rangle$
i.e. $\vert x\rangle\vert y\rangle\rightarrow (1-x)\vert x\rangle\vert y\rangle + x\vert x\rangle\vert 1-y\rangle$

Let the oscillators corresponding to the $\vert x\rangle$ qubit be denoted
by $a,a^{\dagger}$ and those corresponding to the $\vert y\rangle$
qubit be $b,b^{\dagger}$.Then  in terms of oscillator states the CNOT transformation reads:
$${\frac {(a_{1}^{\dagger})^{n_{1}} (a_{2}^{\dagger})^{n_{2}}} {{([n_{1}]![n_{2}]!)^{1/2}}}}
\vert\phi\rangle_{a}
{\frac {(b_{1}^{\dagger})^{k_{1}} (b_{2}^{\dagger})^{k_{2}}} {{([k_{1}]![k_{2}]!)^{1/2}}}}
\vert\phi\rangle_{b}$$
$$\rightarrow                                     
(1-n_{1}){\frac {(a_{1}^{\dagger})^{n_{1}} (a_{2}^{\dagger})^{n_{2}}} {([n_{1}]![n_{2}]!)^{1/2}}}
 \vert\phi\rangle_{a} {\frac {(b_{1}^{\dagger})^{k_{1}} (b_{2}^{\dagger})^{k_{2}}} {([k_{1}]![k_{2}]!)^{1/2}}}
\vert\phi\rangle_{b}$$
$$+n_{1}{\frac {(a_{1}^{\dagger})^{n_{1}} (a_{2}^{\dagger})^{n_{2}}} {([n_{1}]![n_{2}]!)^{1/2}}}
\vert\phi\rangle_{a}
{\frac {(b_{1}^{\dagger})^{k_{2}} (b_{2}^{\dagger})^{k_{1}}} {([k_{2}]![k_{1}]!)^{1/2}}}
\vert\phi\rangle_{b}\eqno(16a)$$
where $n_{1}, n_{2}$ and $k_{1}, k_{2}$ are the eigenvalues of the number operators corresponding 
to the respective oscillators  with $n_{1}+n_{2}=1 , k_{1}+k_{2}=1$ and $\vert\phi\rangle_{a}$,
$\vert\phi\rangle_{b}$ denote the ground states corresponding to oscillators $a_{1,2}$ and $b_{1,2}$ respectively.
Writing,  
$${\frac {(a_{1}^{\dagger})^{n_{1}} (a_{2}^{\dagger})^{n_{2}}} {{([n_{1}]![n_{2}]!)^{1/2}}}}
\vert\phi\rangle_{a}=\vert\eta_{1}\rangle;
{\frac {(b_{1}^{\dagger})^{k_{1}} (b_{2}^{\dagger})^{k_{2}}} {{([k_{1}]![k_{2}]!)^{1/2}}}}
\vert\phi\rangle_{b}=\vert\eta_{2}\rangle;
{\frac {(b_{1}^{\dagger})^{k_{2}} (b_{2}^{\dagger})^{k_{1}}} {{([k_{2}]![k_{1}]!)^{1/2}}}}
\vert\phi\rangle_{b}=\vert -\eta_{2}\rangle$$
the equation $(16a)$ for the CNOT transformation looks like:
$$\vert\eta_{1}\rangle\vert\eta_{2}\rangle\rightarrow (1-n_{1})\vert\eta_{1}\rangle\vert\eta_{2}\rangle
 + n_{1}\vert\eta_{1}\rangle\vert -\eta_{}\rangle\eqno(16b)$$
In all subsequent discussions we shall use this form $(16b)$. However, for completeness, we note that the CNOT 
transformation $(16a)$ can also be written in the alternative notation as 
$$\vert n_{1}- {1\over 2}\rangle_{a} \vert k_{1}- {1\over 2}\rangle_{b}\rightarrow (1-n_{1})\vert n_{1}-{1\over 2}\rangle_{a}
\vert k_{1}-{1\over 2}>_{b} + n_{1}\vert n_{1}- {1\over 2}\rangle_{a}\vert {1\over 2}-k_{1}\rangle_{b}\eqno(16b)$$
i.e.
$$\vert -{1\over 2}\rangle_{a}\vert -{1\over 2}\rangle_{b}\rightarrow\vert -{1\over 2}>_{a}\vert -{1\over 2}\rangle_{b}\enskip;
\enskip\vert -{1\over 2}\rangle_{a}\vert {1\over 2}\rangle_{b}\rightarrow \vert -{1\over 2}\rangle_{a}\vert {1\over 2}\rangle_{b}$$
$$\enskip\enskip\vert {1\over 2}>_{a}\vert -{1\over 2}\rangle_{b}\rightarrow \vert {1\over 2}\rangle_{a}\vert {1\over 2}\rangle_{b}\enskip;
\enskip\vert {1\over 2}\rangle_{a}\vert {1\over 2}>_{b}\rightarrow \vert {1\over 2}\rangle_{a}\vert -{1\over 2}\rangle_{b}$$
For deformed qubits the CNOT transformation will be 
$$\vert x\rangle_{q}\vert y>_{q}\rightarrow (1-x)\vert x>_{q}\vert y\rangle_{q}
 + x\vert x\rangle_{q}\vert 1-y\rangle_{q}\eqno(17a)$$
or in terms of deformed oscillators :
$${\frac {(a_{1q}^{\dagger})^{n_{1}} (a_{2q}^{\dagger})^{n_{2}}} {{([n_{1}]![n_{2}]!)^{1/2}}}}
\vert\phi\rangle_{aq}
{\frac {(b_{1q}^{\dagger})^{k_{1}} (b_{2q}^{\dagger})^{k_{2}}} {{([k_{1}]![k_{2}]!)^{1/2}}}}
\vert\phi\rangle_{bq}$$
$$\rightarrow                                     
(1-n_{1}){\frac {(a_{1q}^{\dagger})^{n_{1}} (a_{2q}^{\dagger})^{n_{2}}} {([n_{1}]![n_{2}]!)^{1/2}}}
 \vert\phi\rangle_{aq} {\frac {(b_{1q}^{\dagger})^{k_{1}} (b_{2q}^{\dagger})^{k_{2}}} {([k_{1}]![k_{2}]!)^{1/2}}}
\vert\phi\rangle_{bq}$$
$$+n_{1}{\frac {(a_{1q}^{\dagger})^{n_{1}} (a_{2q}^{\dagger})^{n_{2}}} {([n_{1}]![n_{2}]!)^{1/2}}}
\vert\phi\rangle_{aq}
{\frac {(b_{1q}^{\dagger})^{k_{2}} (b_{2q}^{\dagger})^{k_{1}}} {([k_{2}]![k_{1}]!)^{1/2}}}
\vert\phi\rangle_{bq}\eqno(17b)$$
As in  Ref.1, the harmonic oscillator 
realisations for the operators $a_{q}, a_{q}^{\dagger}$ and $b_{q}, b_{q}^{\dagger}$ respectively 
are written in terms of the two functions $F$ and $G$ as$^{11,12}$: 
$$a_{1q}^{\dagger}=F(\hat N,q)a_{1}^{\dagger}\enskip \enskip ; 
\enskip\enskip a_{2q}^{\dagger}=F(1-\hat N,q)a_{2}^{\dagger}\eqno(18a)$$
$$b_{1q}^{\dagger}=G(\hat K,q)b_{1}^{\dagger}\enskip\enskip ; 
\enskip\enskip b_{2q}^{\dagger}=G(1-\hat K,q)b_{2}^{\dagger}\eqno(18b)$$
$\hat N$ and $\hat K$ are the respective number operators with eigenvalues $\hat n$ and $\hat k$ and
$$F(\hat N, q)
=\sqrt{{\frac {q^{\hat N} \psi_{1} - q^{-\hat N}\psi_{2}} {{\hat N (q-q^{-1})}}}}\enskip,\enskip
G(\hat K, q)
=\sqrt{{\frac {q^{\hat K} \beta_{1} - q^{-\hat K}\beta_{2}} {{\hat K (q-q^{-1})}}}}\enskip,\enskip
\eqno(19)$$
Using these expressions in $(17b)$ (and relabeling $n_{1}$ as $n$ and $k_{1}$ as $k$ etc.)
and suppressing the $q$ dependence in $F$ and $G$ to avoid cumbersome notation one gets
$$F^{n}(\hat N)F(1-\hat N + n)^{1-n} 
{(a_{1}^{\dagger})^{n}(a_{2}^{\dagger})^{1-n}\over ([n]![1-n]!)^{{1\over 2}}}\vert\phi\rangle_{aq}
G^{k}(\hat K)G(1-\hat K + k)^{1-k} 
{(b_{1}^{\dagger})^{k}(b_{2}^{\dagger})^{1-k}\over ([k]![1-k]!)^{{1\over 2}}}\vert\phi\rangle_{bq}$$
$\longrightarrow$
$$(1-n)F^{n}(\hat N,q)F(1-\hat N + n)^{1-n}
{(a_{1}^{\dagger})^{n}(a_{2}^{\dagger})^{1-n}\over ([n]![1-n]!)^{{1\over 2}}}\vert\phi\rangle_{aq}
G^{k}(\hat K,q)G(1-\hat K + k)^{1-k} 
{(b_{1}^{\dagger})^{k}(b_{2}^{\dagger})^{1-k}\over ([k]![1-k]!)^{{1\over 2}}}\vert\phi\rangle_{bq}$$
$$+n F^{n}(\hat N,q)F(1-\hat N + n)^{1-n}
{(a_{1}^{\dagger})^{n}(a_{2}^{\dagger})^{1-n}\over ([n]![1-n]!)^{{1\over 2}}}\vert\phi\rangle_{aq}
G^{1-k}(\hat K,q)G(1-\hat K + k)^{k}
{(b_{1}^{\dagger})^{1-k}(b_{2}^{\dagger})^{k}\over ([1-k]![k]!)^{{1\over 2}}}\vert\phi\rangle_{bq}\eqno(20a)$$
Denoting
$${\frac {(a_{1}^{\dagger})^{n_{1}} (a_{2}^{\dagger})^{n_{2}}} {{([n_{1}]![n_{2}]!)^{1/2}}}}
\vert\phi\rangle_{aq}=\vert\beta_{1q}\rangle;
{\frac {(b_{1}^{\dagger})^{k_{1}} (b_{2}^{\dagger})^{k_{2}}} {{([k_{1}]![k_{2}]!)^{1/2}}}}
\vert\phi\rangle_{bq}=\vert\beta_{2q}\rangle;
{\frac {(b_{1}^{\dagger})^{k_{2}} (b_{2}^{\dagger})^{k_{1}}} {{([k_{2}]![k_{1}]!)^{1/2}}}}
\vert\phi\rangle_{bq}=\vert -\beta_{2q}\rangle$$
$$F^{n}(\hat N)F(1-\hat N + n)^{1-n}=A$$
$$G^{k}(\hat K)G(1-\hat K + k)^{1-k}=B$$
$$G^{1-k}(\hat K)G(1-\hat K + k)^{k}=B'$$
the equation $(20a)$ becomes
$$A\vert\beta_{1q}\rangle B\vert\beta_{2q}\rangle\longrightarrow 
(1-n)A\vert\beta_{1q}\rangle B\vert\beta_{2q}\rangle + n A \vert\beta_{1q}\rangle B'\vert -\beta_{2q}\rangle\eqno(20b)$$
Multiplying both sides of $(20b)$ by $(AB)^{-1}$ gives 
$$\vert\beta_{1q}\rangle\vert\beta_{2q}\rangle\longrightarrow (1-n)\vert\beta_{1q}\rangle\vert\beta_{2q}\rangle 
+ n\vert\beta_{1q}\rangle B^{-1}B'\vert -\beta_{2q}\rangle\eqno(20c)$$
Note that with respect to the states ,$(20c)$ will be indistinguishable from the usual CNOT 
transformation $(16b)$ if $B^{-1}B'=I$ (the identity operator) i.e. $B=B'$ or 
$$\Biggl({q^{\hat K}\beta_{1}-q^{-\hat K}\beta_{2}\over\hat K (q-q^{-1})}\Biggr)^{{k\over 2}}
\Biggl({q^{1-\hat K+k}\beta_{1}-q^{-(1-\hat K+k)}\beta_{2}\over (1-\hat K+k)(q-q^{-1})}\Biggr)^{{1-k\over 2}}$$
$$=\Biggl({q^{1-\hat K}\beta_{1}-q^{-(1-\hat K)}\beta_{2}\over (1-\hat K) (q-q^{-1})}\Biggr)^{{1-k\over 2}}
\Biggl({q^{\hat K-1+k}\beta_{1}-q^{-(\hat K-1+k)}\beta_{2}\over (\hat K-1+k) (q-q^{-1})}\Biggr)^{{k\over 2}}
\eqno(21)$$
Equation $(21)$ is an identity for both $k=0$ and $k=1$.
Therefore the condition $B=B'$ is always realisable in the domain of $k$.
So the two qubit CNOT gate can be realised
with q-deformed oscillators. Hence all the gates required for universality can also be realised 
with q-deformed oscillators. This implies that any quantum logic gate can be realised with q-deformed oscillators.
Thus quantum computation has an alternative formalism.

{\bf 5. The possible states}

There are two possibilities as regards the arbitrary functions $\psi_{1,2} , \beta_{1,2}$.

{\bf Case:1}
All of them are unity and hence $N=\hat N$ and similarly $K=\hat K$.
So $(2a)$ just relates the opertors $a, a^{\dagger}$ with
$a_{q}, a_{q}^{\dagger}$.A similar argument holds for the operators $b,b^{\dagger}$
and $b_{q},b_{q}^{\dagger}$. Also from $(2b)$ we then have $N=\hat N$ and $K=\hat K$.
This means that at the occupation number level the deformed states cannot be distinguished 
from the undeformed states and we are in the realm of standard quantum computation.
We denote eigenvalues of the number operators for deformed oscillators 
in Case I by $n,k$  ($\hat n,\hat k$ still correspond to undeformed oscillators);
the states in Case I by $\vert \rangle_{I}$.Then relabel $n_{1}$ by $n$ etc. we have for Case I
$$\vert n-1/2\rangle_{I}\vert k-1/2\rangle_{I}
=\vert n\rangle_{Ia_{1}}\vert 1- n\rangle_{Ia_{2}}\vert k\rangle_{Ib_{1}}\vert 1-k\rangle_{Ib_{2}}$$
$$=\vert\hat n\rangle_{Ia_{1}}\vert 1-\hat n\rangle_{Ia_{2}}\vert\hat k\rangle_{Ib_{1}}\vert 1-\hat k\rangle_{Ib_{2}}\eqno(22a)$$
where  $n=0,1; k=0,1$ and $n=\hat n; k=\hat k$. Note that all states have $j={1\over 2}$.The  possible states are:
$$\vert 00\rangle_{I}=\vert -{1\over 2}\enskip -{1\over 2}\rangle_{I}=\vert -{1\over 2}\rangle_{Ia}\vert -{1\over 2}\rangle_{Ib}
=\vert \tilde 0\rangle_{Ia_{1}}\vert\tilde 1\rangle_{Ia_{2}}\vert\tilde 0\rangle_{Ib_{1}}\vert\tilde 1\rangle_{Ib_{2}}\eqno(22b)$$
$$\vert 01\rangle_{I}=\vert -{1\over 2}\enskip {1\over 2}\rangle_{I}=\vert -{1\over 2}\rangle_{Ia}\vert {1\over 2}\rangle_{Ib}
=\vert \tilde 0\rangle_{Ia_{1}}\vert\tilde 1\rangle_{Ia_{2}}\vert\tilde 1\rangle_{Ib_{1}}\vert\tilde 0\rangle_{Ib_{2}}\eqno(22c)$$
$$\vert 10\rangle_{I}=\vert {1\over 2}\enskip -{1\over 2}\rangle_{I}=\vert {1\over 2}\rangle_{Ia}\vert -{1\over 2}\rangle_{Ib}
=\vert \tilde 1\rangle_{Ia_{1}}\vert\tilde 0\rangle_{Ia_{2}}\vert\tilde 0\rangle_{Ib_{1}}\vert\tilde 1\rangle_{Ib_{2}}\eqno(22d)$$
$$\vert 11\rangle_{I}=\vert {1\over 2}\enskip {1\over 2}\rangle_{I}=\vert {1\over 2}\rangle_{Ia}\vert {1\over 2}\rangle_{Ib}
=\vert \tilde 1\rangle_{Ia_{1}}\vert\tilde 0\rangle_{Ia_{2}}\vert\tilde 1\rangle_{Ib_{1}}\vert\tilde 0\rangle_{Ib_{2}}\eqno(22e)$$

{\bf Case:II}

We have a general scenario if the arbitrary functions $\psi_{i}(q),\beta_{i}(q) i=1,2$ are
not all equal to unity. As these are arbitrary, let us choose $\psi_{1}=\psi_{2}=\psi , \beta_{1}=\beta_{2}=\beta$.
Then $N=\hat N - (1/s)~ln~ \psi(q)$ [$(2b)$];
and $K=\hat K - (1/s)~ln~\beta(q)$.

Hence states labelled by the  occupation number are different as the eigenvalues of the
number operator of  standard oscillator states and the
eigenvalues of the number operator of deformed oscillator states are now related by
$n=\hat n - (1/s)~ ln ~\psi(q)$ ; $k=\hat k - (1/s)~ ln ~\beta(q)$.
This would show up in the Jordan-Schwinger construction.
We denote eigenvalues of the number operators for deformed oscillators 
in Case II by $n',k'$ and the states by $\vert >_{II}$.
So
$$\vert n'-1/2\rangle_{II}\vert k'-1/2\rangle_{II}$$
$$=\vert n'\rangle_{IIa_{1}}\vert 1-n'\rangle_{IIa_{2}}\vert k'\rangle_{IIb_{1}}\vert 1-k'\rangle_{IIb_{2}}$$
$$=\vert\hat n - (1/s) ln \psi\rangle_{IIa_{1}}\vert 1-\hat n + (1/s) ln\psi\rangle_{IIa_{2}}
\vert\hat k - (1/s) ln \beta\rangle_{IIb_{1}}\vert 1-\hat k + (1/s) ln\beta\rangle_{IIb_{2}}
\eqno(23a)$$
All possible states are :
$$\vert 00\rangle_{II}=\vert -{1\over 2}\enskip -{1\over 2}\rangle_{II}=\vert -{1\over 2}\rangle_{IIa}\vert -{1\over 2}\rangle_{IIb}
=\vert \tilde 0\rangle_{IIa_{1}}\vert\tilde 1\rangle_{IIa_{2}}\vert\tilde 0\rangle_{IIb_{1}}\vert\tilde 1\rangle_{IIb_{2}}\eqno(23b)$$
$$\vert 01\rangle_{II}=\vert -{1\over 2}\enskip {1\over 2}\rangle_{II}=\vert -{1\over 2}\rangle_{IIa}\vert {1\over 2}\rangle_{IIb}
=\vert \tilde 0\rangle_{IIa_{1}}\vert\tilde 1\rangle_{IIa_{2}}\vert\tilde 1\rangle_{IIb_{1}}\vert\tilde 0\rangle_{IIb_{2}}\eqno(23c)$$
$$\vert 10\rangle_{II}=\vert {1\over 2}\enskip -{1\over 2}\rangle_{II}=\vert {1\over 2}\rangle_{IIa}\vert -{1\over 2}\rangle_{IIb}
=\vert \tilde 1\rangle_{IIa_{1}}\vert\tilde 0\rangle_{IIa_{2}}\vert\tilde 0\rangle_{IIb_{1}}\vert\tilde 1\rangle_{IIb_{2}}\eqno(23d)$$
$$\vert 11\rangle_{II}=\vert {1\over 2}\enskip {1\over 2}\rangle_{II}=\vert {1\over 2}\rangle_{IIa}\vert {1\over 2}\rangle_{IIb}
=\vert \tilde 1\rangle_{IIa_{1}}\vert\tilde 0\rangle_{IIa_{2}}\vert\tilde 1\rangle_{IIb_{1}}\vert\tilde 0\rangle_{IIb_{2}}\eqno(23e)$$

Consistency demands the following interpretations:

(1)For $(23b)$, $\psi=q^{\hat n}$; $\beta=q^{\hat k}$ i.e. the qubit state  
$\vert\tilde 0\rangle_{IIa_{1}}\vert\tilde 1\rangle_{IIa_{2}}$ corresponds to an oscillator occupation number $\hat n>0$ while
$\vert\tilde 0\rangle_{IIb_{1}}\vert\tilde 1\rangle_{IIb_{2}}$
corresponds to an oscillator occupation number $\hat k>0$.

(2)For $(23c)$, $\psi=q^{\hat n}$; $\beta=q^{\hat k-1}$ i.e. the qubit state  
$\vert\tilde 0\rangle_{IIa_{1}}\vert\tilde 1\rangle_{IIa_{2}}$ corresponds to an oscillator occupation number $\hat n>0$ while 
$\vert\tilde 1\rangle_{IIb_{1}}\vert\tilde 0\rangle_{IIb_{2}}$
corresponds to an oscillator occupation number $\hat k>1$.

(3)For $(23d)$, $\psi=q^{\hat n-1}$; $\beta=q^{\hat k}$ i.e. the qubit state  
$\vert\tilde 1\rangle_{IIa_{1}}\vert\tilde 0\rangle_{IIa_{2}}$ corresponds to an oscillator occupation number $\hat n>1$ while 
$\vert\tilde 0\rangle_{IIb_{1}}\vert\tilde 1\rangle_{IIb_{2}}$
corresponds to an oscillator occupation number $\hat k>0$.

(4)For $(23e)$, $\psi=q^{\hat n-1}$; $\beta=q^{\hat k-1}$ i.e. the qubit state 
$\vert\tilde 1\rangle_{IIa_{1}}\vert\tilde 0\rangle_{IIa_{2}}$ corresponds to an oscillator occupation number $\hat n>1$ while 
$\vert\tilde 1\rangle_{IIb_{1}}\vert\tilde 0\rangle_{IIb_{2}}$
corresponds to an oscillator occupation number $\hat k>1$.

Therefore we always have $\hat n > n', \hat k > k' $. $\psi(q) , \beta(q)$ cannot be unity
(i.e. $\hat n, \hat k$ cannot be zero) because then we will have $n'=\hat n, k'=\hat k$
i.e.Case I. So the deformed states in Case II can be related
to harmonic oscillator states with occupation numbers greater than zero.

Denote the $F$ and $G$ functions corresponding to the two possibilities by 
$F_{I} , G_{I}$ and $F_{II} , G_{II}$.Then  
$$F_{I}(\hat N, q)
=\Biggl({q^{\hat N} - q^{-\hat N}\over \hat N (q-q^{-1})}\Biggr)^{1\over 2}\enskip;\enskip
G_{I}(\hat K, q)
=\Biggl({q^{\hat K} - q^{-\hat K}\over \hat K (q-q^{-1})}\Biggr)^{1\over 2}\eqno(24)$$
$$F_{II}(\hat N, q)
=\Biggl({q^{\hat N}\psi_{1} - q^{-\hat N}\psi_{2}\over\hat N (q-q^{-1})}\Biggl)^{1\over 2}\enskip;\enskip
G_{II}(\hat K, q)
=\Biggl({q^{\hat K}\beta_{1} - q^{-\hat K}\beta_{2}\over\hat K (q-q^{-1})}\Biggr)^{1\over 2}\eqno(25)$$
where we have labelled the arbitrary functions by $\psi_{1,2}$ and $\beta_{1,2}$.
Now the properties of the operators $F$ and $G$ have to be understood in terms of their 
eigenvalues.Then the ratio of the eigenvalues of $F_{II}$ and $F_{I}$ is (choosing $\psi_{1}=\psi_{2}=\psi$
and $\beta_{1}=\beta_{2}=\beta$)
$${Eigenvalue\enskip\enskip of\enskip\enskip F_{II}\over Eigenvalue\enskip\enskip of\enskip\enskip F_{I}}
=\Biggl({q^{2\hat n}\psi_{1}(q)-\psi_{2}(q)\over q^{2\hat n}-1}\Biggr)^{1/2}=\psi ^{1\over 2}(q)\eqno(26a)$$
So we may write 
$$F_{II}=\psi ^{1\over 2}(q)F_{I}\eqno(26b)$$
Similarly
$${Eigenvalue\enskip\enskip of\enskip\enskip G_{II}\over Eigenvalue\enskip\enskip of\enskip\enskip G_{I}}
=\Biggl({q^{2\hat k}\beta_{1}(q)-\beta_{2}(q)\over q^{2\hat k}-1}\Biggr)^{1/2}=\beta ^{1\over 2}(q)\eqno(27a)$$
So we may write 
$$G_{II}= \beta ^{1\over 2}(q)G_{I}\eqno(27b)$$
Thus
$$\vert n',k'\rangle_{II}={\frac {(F_{II}a_{1}^{\dagger})^{n'_{1}} (F_{II}a_{2}^{\dagger})^{n'_{2}}} {{([n'_{1}]![n'_{2}]!)^{1/2}}}}
\vert\phi\rangle_{aq}
{\frac {(G_{II}b_{1}^{\dagger})^{k'_{1}} (G_{II}b_{2}^{\dagger})^{k'_{2}}} {{([k'_{1}]![k'_{2}]!)^{1/2}}}}
\vert\phi\rangle_{bq}$$
$$={\frac {(\psi ^{1\over 2}F_{I}(\hat N)a_{1}^{\dagger})^{n'}
(\psi ^{1\over 2}F_{I}(1-\hat N)a_{2}^{\dagger})^{1-n'}} {{([n']![1-n']!)^{1/2}}}}
\vert\phi\rangle_{aq}$$
$${\frac {(\beta ^{1\over 2}G_{I}(\hat K)b_{1}^{\dagger})^{k'} 
(\beta ^{1\over 2}G_{I}(1-\hat K)b_{2}^{\dagger})^{1-k'}} {{([k']![1-k']!)^{1/2}}}}
\vert\phi\rangle_{bq}$$
$$=\psi ^{n'\over 2}\psi ^{{1-n'\over 2}}\beta ^{k'\over 2}\beta^ {{1-k'\over 2}}\vert n,k\rangle_{I}
=\psi ^{1\over 2}\beta ^{1\over 2}\vert n,k\rangle_{I}\eqno(28)$$
Therefore
$${_{II}\langle n',k'\vert n',k'\rangle_{II}\over_{I}\langle n,k\vert n,k\rangle_{I}}$$
$$=\psi^{1\over 2}(q)\beta ^{1\over 2}(q)\eqno(29)$$

So the right hand side of $(29)$ is a function of $q$ only.
For 
$\psi(q)=1$ and $\beta(q)=1$, one cannot distinguish between the two cases at the 
level of the scalar products between states. However, if the arbitrary functions 
are not unity then these scalar products are distinct from each other and this 
might be useful at the level of experimental realisations or consequences.
Moreover, note that one can choose the arbitrary function $\psi(q)$ for the control 
qubit to be the same as that in the Hadamard, Phase shift and NOT gates. As the CNOT gate 
is a two qubit gate, a different function $\beta(q)$ is taken for the target qubit.
So with two arbitrary functions all the three gates required for universality 
can be constructed with q-deformed qubits

{\bf 6.Conclusion}

Therefore quantum computation admits of an alternative formalism where q-deformed oscillator 
states can be used to construct qubits. This has been established here with the realisation 
of the CNOT quantum logic gate with q-deformed oscillators. Thus this realisation is possible 
for all the quantum logic gates required for  universality. Hence {\it all quantum logic 
gates can be realised with q-deformed qubits}. The existence of additional parameters will enable 
comparison between different experimental scenarios using the usual scheme and the 
alternate scheme. This requires further study. 

Another aspect where the new formalism might prove useful is the realm of quantum error 
correction. The conventional way to correct computer error is to use redundancy.
More than one element is used to denote the same bit.For example, consider two atoms 
$A$ and $B$ and use the doublet $AB$ to store the same bit (of information). Thus 
one has the state $\vert 0 0\rangle=\vert 0\rangle\vert 0\rangle$ (or
the state $\vert 1 1\rangle=\vert 1\rangle\vert 1\rangle$).An error changes the state 
of only one atom, i.e. one now has any one of the states $\vert 01\rangle$, $\vert 10\rangle$ 
instead of $\vert 00\rangle$ (or the states $\vert 10\rangle$, $\vert 01\rangle$ instead of 
$\vert 11\rangle$). For type II states, $\vert\rangle_{II}$, we have $n'=\hat n -{1\over s}ln\psi$. 
It is simple to check that $\psi=q^{\hat n}$ for $n'=0$ and $\psi=q^{\hat n -1}$ for $n'=1$. Therefore 
occurrence of an error is reflected in the change of the arbitrary function and consequently 
in the matrix elements $_{II}\langle n'\vert n'\rangle_{II}$.
So the arbitrary functions provide an additional leverage and can possibly 
be used in detecting and regulating errors. This also requires further 
investigation.

\end{document}